\documentstyle[12pt]{article}

\begin{document}
\title{Part of the D - dimensional Spiked harmonic oscillator spectra}
\author{ Omar Mustafa and Maen Odeh\\
 Department of Physics, Eastern Mediterranean University\\
 G. Magusa, North Cyprus, Mersin 10 - Turkey\\
 email: omustafa.as@mozart.emu.edu.tr\\
\date{}\\}
\maketitle
\begin{abstract}
{\small The pseudoperturbative shifted - $l$ expansion technique PSLET
[5,20] is generalized for states with arbitrary number of nodal zeros.
Interdimensional degeneracies, emerging from the isomorphism between
angular momentum and dimensionality of the central force Schr\"odinger
equation, are used to construct part of the $D$ - dimensional spiked
harmonic oscillator bound - state spectra. PSLET results are found to
compare excellently
with those from direct numerical integration and generalized variational
methods [1,2].}
\end{abstract}
\newpage

\section{Introduction}

The simplest model of realistic interaction potentials in atomic, molecular,
and nuclear physics is provided by the spiked harmonic oscillator
\begin{equation}
V(q)=c_1 q^2+c_2 q^{-b}~~,~~~~ c_1, c_2, b>0~~,~~~~q\in(0, \infty).
\end{equation}\\
The construction of its bound - states has attracted attention over the
last few years [1-9]. It is an interesting model not only because of being
a singular potential representing a repulsive core in realistic interactions,
but also because of its intrinsic properties in view of mathematical physics
[10-16]. However, most of the studies on this model potential (1) were
devoted to one spatial dimension ( $1D$, the hyperquantum limit in view
of Herschbach [17,18]). It was just very recently, to the best of our
knowledge, that Hall and Saad have generalized their variational
analysis [1a] and smooth transformation [2] methods, VAM and STM, respectively, to the
D - dimensional case and studied its bound - states. They have also used
direct numerical integration (DNI) for comparison purposes. It is therefore
interesting to carry out systematic studies of the bound - state spectra
generated by this interesting class of singular potentials (1).

On the other hand, results from exactly solvable potentials ( an interesting
field of mathematical physics in itself) are essential ingredients for
the description of realistic physical problems [1-5,19].
The solutions of these can be used in
perturbation and pseudoperturbation theories, or they can be combined with
numerical calculations. Nevertheless, in the simplest case, analytical
calculations can aid numerical studies in areas where numerical techniques
might not be safely controlled. For example, when bound - state wave
functions with arbitrary nodal zeros are required for certain singular
potentials (a next level of complexity), analytical solutions can supply
a basis for numerical calculations. Moreover, in many problems the
Hamiltonian does not contain any physical parameter suitable for a
perturbation expansion treatment. More often, the Hamiltonian contains
physical parameters, but, typically, zeroth - order solutions for special
values of these are not tractable or good starting approximations. One
would therefore resort to variational calculations [1], pseudoperturbation
expansions ( artificial in nature) [5,18-26], etc.

Recently, we have introduced a pseudoperturbative shifted - $l$ ( $l$ is the
angular momentum quantum number) expansion technique ( PSLET) to solve for
nodeless states of Schr\"odinger equation. It simply consists of using
$1/\bar{l}$ as a pseudoperturbation parameter, where $\bar{l}=l-\beta$
and $\beta$ is a suitable shift. The shift $\beta$ is vital for it removes
the poles  that would emerge, at lowest orbital states with
$l$=0, in our proposed expansions below. Our analytical, or often
semianalytical, methodical proposal PSLET has been successfully applied
to quasi - relativistic harmonic oscillator [20], spiked harmonic oscillator
[5], anharmonic oscillators [21], and to the two - dimensional ({\em
flatland}, in view of Godson and L\'opez - Cabrera in [17]) hydrogenic atom
in an arbitrary magnetic field [22].

Encouraged by its satisfactory performance in handling nodeless states,
we generalize PSLET recipe ( in section 2) for states with arbitrary number
of nodal zeros, $k \geq 0$. Moreover, in the underlying "radical" time -
independent radial Schr\"odinger equation, in $\hbar=m=1$ units,\\
\begin{equation}
\left[-\frac{1}{2}\frac{d^{2}}{dq^{2}}+\frac{l(l+1)}{2q^{2}}+V(q)\right]
\Psi_{k,l}(q)=E_{k,l}\Psi_{k,l}(q),
\end{equation}\\
the isomorphism between orbital angular momentum $l$ and  dimensionality
$D$ invites interdimensional degeneracies to obtain [17].  Which, in effect,
allows us to generate the ladder of excited states for any given $k$ and
nonzero $l$ from the $l$=0 result, with that $k$, by the transcription
$D \longrightarrow D+2l$. That is, if $E_{k,l}(D)$ is the eigenvalue in $D$
- dimensions, then\\
\begin{equation}
E_{k,l}(2) \equiv E_{k,l-1}(4) \equiv  \cdots 
\equiv E_{k,1}(2l) \equiv E_{k,0}(2l+2)
\end{equation}\\
for even $D$, and\\
\begin{equation}
E_{k,l}(3) \equiv E_{k,l-1}(5)  \equiv \cdots \equiv
E_{k,1}(2l+1) \equiv E_{k,0}(2l+3)
\end{equation}\\
for odd $D$. For more details the reader may refer to ref.s [17,18,27].
We therefore calculate, in section 3, the energies for $D=2$ and $D=3$
spiked harmonic oscillators, for a given number of nodes $k$ and different
values of $l$, and construct part of its $D$ - dimensional bound - state
spectra. We compare our results with those reported by Hall and Saad via
generalized variational analysis VAM, and direct
numerical integration DNI methods [1,2]. Section 4 is devoted for concluding
remarks.

\section{The generalization of PSLET}

With the shifted angular momentum, equation (2) reads\\
\begin{equation}
\left\{-\frac{1}{2}\frac{d^{2}}{dq^{2}}+\frac{\bar{l}^{2}+(2\beta+1)\bar{l}
+\beta(\beta+1)}{2q^{2}}+\frac{\bar{l}^2}{Q}V(q) \right\}
\Psi_{k,l} (q)=E_{k,l}\Psi_{k,l}(q),
\end{equation}\\
where Q is a constant that scales the potential $V(q)$ at large - $l_D$ limit
( the pseudoclassical limit [17]) and is set, for any specific choice of
$l_D$ and $k$, equal to $\bar{l}^2$ at the end of the calculations. Here
$l_D=l+(D-3)/2$, to incorporate the interdimensional degeneracies associated
with the isomorphism between angular momentum and dimensionality $D$. Hence,
$\bar{l} \longrightarrow \bar{l}=l_D - \beta$ through out this paper. Next,
we shift the origin of the coordinate system through
$x=\bar{l}^{1/2}(q-q_{o})/q_{o}$, where $q_o$ is currently an arbitrary
point to be determined below. Expansions about this point
(see Appendix for more details), $x=0$ (i.e. $q=q_o$),
obviously localize the problem at an
arbitrary point $q_o$ and the derivatives, in effect, contain
information not only at $q_o$ but also at any point on $q$-axis, in
accordance with Taylor's theorem. It is then convenient to expand $E_{k,l}$
 as\\
\begin{equation}
E_{k,l}=\sum^{\infty}_{n=-2}E_{k,l}^{(n)}~\bar{l}^{-n}.
\end{equation}\\
Equation (5) thus becomes\\
\begin{equation}
\left[-\frac{1}{2}\frac{d^2}{dx^2} + \sum^{\infty}_{n=0} v^{(n)}
\bar{l}^{-n/2}\right]\Psi_{k,l} (x)= 
\left[\sum^{\infty}_{n=1} q_o^2 E_{k,l}^{(n-1)}
\bar{l}^{-n} \right] \Psi_{k,l}(x).
\end{equation}\\

Up to this point, one would conclude that the above procedure is nothing
but an imitation of the eminent shifted large-N expansion (SLNT)
[25,26,28-30]. However, because of the limited capability of SLNT
in handling large-order corrections via the standard Rayleigh-Schr\"odinger
perturbation theory, only low-order corrections have been reported,
sacrificing in effect its preciseness. Therefore, one should seek
for an alternative and proceed by setting the wave functions with any number
of nodes as \\
\begin{equation}
\Psi_{k,l}(x(q)) = F_{k,l}(x)~ exp(U_{k,l}(x)).
\end{equation}\\
In turn, equation (7) readily transforms into the
following Riccati equation:\\
\begin{eqnarray}
&&F_{k,l}(x)\left[-\frac{1}{2}\left( U_{k,l}^{''}(x)+U_{k,l}^{'}(x)
U_{k,l}^{'}(x)\right)
+\sum^{\infty}_{n=0} v^{(n)}(x) \bar{l}^{-n/2} \right. \nonumber\\
&&\left. -\sum^{\infty}_{n=1} q_o^2 E_{k,l}^{(n-1)} \bar{l}^{-n} \right]
-F_{k,l}^{'}(x)U_{k,l}^{'}(x)-\frac{1}{2}F_{k,l}^{''}(x)=0,
\end{eqnarray}\\
where the primes denote derivatives with respect to $x$. It is
evident that this equation admits solution of the form \\
\begin{equation}
U_{k,l}^{'}(x)=\sum^{\infty}_{n=0} U_{k}^{(n)}(x)~~\bar{l}^{-n/2}
+\sum^{\infty}_{n=0} G_{k}^{(n)}(x)~~\bar{l}^{-(n+1)/2},
\end{equation}\\
\begin{equation}
F_{k,l}(x)=x^k +\sum^{\infty}_{n=0}\sum^{k-1}_{p=0}
a_{p,k}^{(n)}~~x^p~~\bar{l}^{-n/2},
\end{equation}\\
where\\
\begin{equation}
U_{k}^{(n)}(x)=\sum^{n+1}_{m=0} D_{m,n,k}~~x^{2m-1} ~~~~;~~~D_{0,n,k}=0,
\end{equation}\\
\begin{equation}
G_{k}^{(n)}(x)=\sum^{n+1}_{m=0} C_{m,n,k}~~x^{2m}.
\end{equation}\\
Substituting equations (10) - (13) into equation (9) implies\\
\begin{eqnarray}
&&F_{k,l}(x)\left[-\frac{1}{2}\sum^{\infty}_{n=0}\left(U_{k}^{(n)^{'}}
\bar{l}^{-n/2}
+ G_{k}^{(n)^{'}} \bar{l}^{-(n+1)/2}\right) \right. \nonumber\\
&-&\left.\frac{1}{2} \sum^{\infty}_{n=0} \sum^{n}_{m=0}
\left( U_{k}^{(m)}U_{k}^{(n-m)} \bar{l}^{-n/2}
+G_{k}^{(m)}G_{k}^{(n-m)} \bar{l}^{-(n+2)/2}
\right. \right.\nonumber\\
&+&\left.\left.2 U_{k}^{(m)}G_{k}^{(n-m)} \bar{l}^{-(n+1)/2}\right)
+\sum^{\infty}_{n=0}v^{(n)} \bar{l}^{-n/2}
-\sum^{\infty}_{n=1} q_o^2 E_{k,l}^{(n-1)} \bar{l}^{-n}\right] \nonumber\\
&-&F_{k,l}^{'}(x)\left[\sum^{\infty}_{n=0}\left(U_{k}^{(n)}\bar{l}^{-n/2}
+ G_{k}^{(n)} \bar{l}^{-(n+1)/2}\right)\right]-\frac{1}{2}F_{k,l}^{''}(x)
=0
\end{eqnarray}\\
The above procedure
obviously reduces to the one described by Mustafa and Odeh [5,20-22], for
$k=0$. Moreover, the solution of equation (14) follows from the uniqueness
of power series representation. Therefore, for a given $k$ we equate the
coefficients of the same powers of $\bar{l}$ and $x$, respectively. 
For example, when $k=1$ one obtains\\
\begin{equation}
D_{1,0,1}=-w,~~~ U_{1}^{(0)}(x) =-~w~x, 
\end{equation}\\
\begin{equation}
C_{1,0,1}=-\frac{B_{3}}{w},~~~~a_{0,1}^{(1)}=-\frac{C_{0,0,1}}{w},
\end{equation}\\
\begin{equation}
C_{0,0,1}=\frac{1}{w}\left(2C_{1,0,1}+2\beta+1\right),
\end{equation}\\
\begin{equation}
D_{2,2,1}=\frac{1}{w}\left(\frac{C_{1,0,1}^2}{2}-B_{4}\right),
\end{equation}\\
\begin{equation}
D_{1,2,1}=\frac{1}{w}\left(\frac{5}{2}~D_{2,2,1}+C_{0,0,1}~C_{1,0,1}
-\frac{3}{2}(2\beta+1)\right),
\end{equation}\\
\begin{equation}
E_{1,l}^{(0)} = \frac{1}{q_o^2}\left(\frac{\beta(\beta+1)}{2}+
a_{0,1}^{(1)}~C_{1,0,1}-\frac{3~D_{1,2,1}}{2}-\frac{C_{0,0,1}^2}{2}\right),
\end{equation}\\
etc. Here, we reported the nonzero coefficients only and give the
definitions of the related parameters in the Appendix. One can then calculate
the energy eigenvalues and eigenfunctions from the knowledge of
$C_{m,n,k}$, $D_{m,n,k}$, and $a_{p,k}^{(n)}$ in a hierarchical manner.
Nevertheless, the procedure just described is suitable for a
software package such as  MAPLE to determine
the energy eigenvalue and eigenfunction corrections up to any order of the
pseudoperturbation series (6). 

Although the energy series, equation (6), could appear
divergent, or, at best, asymptotic for small $\bar{l}$, one can still 
calculate the eigenenergies to a very good accuracy by forming the 
sophisticated [N,M] Pad\'e approximation [24]\\
\begin{center}
$P_{N}^{M}(1/\bar{l})=(P_0+P_1/\bar{l}+\cdots+P_M/\bar{l}^M)/
(1+q_1/\bar{l}+\cdots+q_N/\bar{l}^N)$
\end{center}
to the energy series (6). The energy series (6) is calculated up to
$E_{k,l}^{(8)}/\bar{l}^8$ by
\begin{equation}
E_{k,l}=\bar{l}^{2}E_{k,l}^{(-2)}+E_{k,l}^{(0)}+\cdots
+E_{k,l}^{(8)}/\bar{l}^8+O(1/\bar{l}^{9}),
\end{equation}\\
and with the $P_{4}^{4}(1/\bar{l})$ Pad\'e approximant it becomes\\
\begin{equation}
E_{k,l}[4,4]=\bar{l}^{2}E_{k,l}^{(-2)}+P_{4}^{4}(1/\bar{l}).
\end{equation}\\
Our recipe is therefore well prescribed.

\section{D - spiked harmonic oscillator spectra}

In this section we consider the spiked harmonic oscillator potential (1)
and illustrate the above mentioned procedure. The substitution of equation
(1) in (45), for $k \geq 0$, implies\\
\begin{equation}
w=\sqrt{\frac{8c_1q_o+bc_2(b-2)q_o^{-(b+1)}}
{2c_1q_o-bc_2q_o^{-(b+1)}}}~~,~~ \beta=-\frac{1}{2}(1+[2k+1]w).
\end{equation}\\
Equation (44), in turn, reads\\
\begin{equation}
l_D+\frac{1}{2}\left(1+[2k+1]\sqrt{\frac{8c_1q_o+bc_2(b-2)q_o^{-(b+1)}}
{2c_1q_o-bc_2q_o^{-(b+1)}}}\right)
=q_o^2 \sqrt{c_1-\frac{bc_2}{2}q_o^{-(b+2)}},
\end{equation}\\
which is explicit in $q_o$. However, in the absence of a closed - form
solution for $q_o$, which is often the case ( hence the notion that PSLET
is often semianalytical), numerical solutions of (24) could resolve this
issue. Once $q_o$ is determined the coefficients $C_{m,n,k}$, $D_{m,n,k}$,
and $a_{p,k}^{(n)}$ are determined in a sequential manner. Hence, the
eigenvalues, equation (21), and eigenfunctions, equations (10)-(13), are
calculated in the same batch for each value of $k$, $D$, $l$, $c_1$, $c_2$,
and $b$.

Table 1 shows PSLET results for the ground - state energies, covering a wide
range of the coupling $c_2$ when $b=2.5$, along with those reported by Hall
and Saad [1a], via a generalized variational analysis and
direct numerical integration  methods. Using the interdimensional
degeneracies, equations (3) and (4), or directly the dimensionality $D$
in $l_D$, we display the energies for $V(q)=(q^2+10/q^{1.9})/2$ in table 2.
Clearly, our results compare excellently with those from direct numerical
integrations. However, it should be noted that in [5] we have calculated
the energy series up to $E_{0,l}^{(4)}/\bar{l}^4$ correction. Therefore,
slight discrepancies obtain between the present results in table 1 and those
reported in table 2 of [5].

Adhering to the implicated wisdom in equations (3) and (4), that the  two - and three - dimensional ( 2D and 3D, respectively) cases are the basic
ingredients of the energy ladder at larger dimensions, we report (in
table 3) the 2D -  and 3D - nodal bound - state energies when the coupling
$c_2=1000$ and $b=0.5, 1,\cdots, 2.5, 3$. The stability of the last three
approximants of the Pad\'e sequence indicate that the results are exact.
For more details on this issue the reader may refer to ref.s [20,24].
Nevertheless, our results $E_{0,0}$ for the 3D - spiked harmonic oscillator
are in exact accord with those from direct numerical integrations [2].
Following the same strategy, we display in table 4 the $k=1$ and 2 nodal
bound - state energies for $V(q)=(q^2+1000/q^{3/2})/2$.
Eventually, the leading term of PSLET, $\bar{l}^2E_{k,l}^{(-2)}$,
turns out to be a good starting approximation. Tables 1,2 and 4 bear
this out.

Moreover, for the spiked harmonic oscillator, with $b=2$, one would
rewrite the effective potential term $(l(l+1)+c_2)/2q^2+q^2/2$ as
$l^{'}(l^{'}+1)/2q^2+q^2/2$ with $l^{'}=-1/2+\sqrt{(l+1/2)^2+c_2}$. For this
particular case, PSLET procedure yields, respectively, $w=2$,
$\beta=-(2k+3/2)$, $\bar{l}=2k+l^{'}+3/2$, $q_o^2=\bar{l}$,
$\bar{l}^2 E_{k,l^{'}}^{(-2)}=2k+l^{'}+3/2$ ( the exact well
known energies),\\
\begin{equation}
E_{k,l^{'}}^{(0)}=E_{k,l^{'}}^{(1)}= \cdots=E_{k,l^{'}}^{(8)}=\cdots
=E_{k,l^{'}}^{(n)}=0,
\end{equation}\\
and when $k=0$, for example,\\
\begin{eqnarray}
U_{0,l^{'}}(x)&=&-\frac{1}{2}\left(y-\frac{y^2}{2}+\frac{y^3}{3}-\frac{y^4}{4} +\frac{
y^5}{5}-\frac{y^6}{6}+\frac{y^7}{7}-\frac{y^8}{8}+\cdots\right)  \nonumber \\
&&+\bar{l}\left(y-\frac{y^2}{2}+\frac{y^3}{3}-\frac{y^4}{4}+\frac{y^5}{5} -
\frac{y^6}{6}+\frac{y^7}{7}-\frac{y^8}{8}+\cdots\right)  \nonumber \\
&&-\bar{l}\frac{y^2}{2}-\bar{l}y,
\end{eqnarray}\\
where $y=x\bar{l}^{-1/2}$. Obviously, the terms in brackets in equation (26)
are the infinite geometric series expansions for $ln(1+y)$. Equation (26)
thus becomes\\
\begin{equation}
U_{0,l^{'}}(x) = ln(1+y)^{-1/2} + ln(1+y)^{\bar{l}} - \bar{l}y - \bar{l}\frac{y^2
}{2}.
\end{equation}\\
Hence Eq.(8) ( with $F_{0,l^{'}}(x)=1$ from (11)) reads\\
\begin{equation}
\Psi_{0,l^{'}} (q) = N_{0,l^{'}}~~q^{l^{'}+1}~~e^{-q^2 /2},
\end{equation}\\
the exact well known solutions [31], where $N_{0,l^{'}}$ are the normalization
constants. Proceeding exactly as above, one could obtain the well known
solutions with $k \geq 1$. However, this already lies far beyond the scope
of our present methodical proposal.

Hall and Saad [1a] have therefore used, indirectly, the transformation of
the angular momentum quantum number and cast the Hamiltonian of the
spiked harmonic oscillator (1) as\\
\begin{equation}
H=-\frac{1}{2}\frac{d^{2}}{dq^{2}}+\frac{l_H(l_H+1)}{2q^{2}}
+\frac{q^2}{2}+\frac{c_2}{2q^b}-\frac{A}{2q^2}.
\end{equation}\\
Where $l_H=-1/2 + \sqrt{(l+1/2)^2+A}$, and A is used as a {\em further
variational refinement} in their generalized variational analysis method.
They found that $A=c_2$ is a {\em good} general estimate for the value of $A$.
Indeed this optimum value of $A$, which reduced substantially the number
of the basis function needed for a given accuracy in [1a], enhances the
convergence and accuracy of approximation methodical recipes. Practically,
it minimizes the effect of the perturbation term $c_2q^{-b}$ over the
harmonic oscillator one (with the irrational quantum number $l^{'}$),
especially for values of $b \longrightarrow  2$. In table 5, the results of
PSLET are obtained using such prescription. They compare excellently with
direct numerical integrations and do not contradict with the upper bounds
from the generalized variational estimates.

\section{Concluding remarks}

We have generalized our pseudoperturbative shifted - $l$ expansion
technique PSLET [5,20-22] for states with arbitrary number of nodal
zeros, $k \geq 0$. Starting with the central force problem,
represented by the radial Schr\"odinger equation, and augmenting the
orbital angular momentum by $l \longrightarrow l_D=l+(D-3)/2$, we have
incorporated interdimensional degeneracies. To test PSLET performance,
we have treated the spiked harmonic oscillator problem in $D$ - dimensions.
and used results from direct numerical integrations and generalized
variational analysis methods [1,2] to compare with. The comparison is readily
satisfactory.

The salient features of the attendant proposal PSLET are in order.

It avoids troublesome questions as those pertaining to the nature
of small parameter expansions, the trend of convergence to the exact
numerical values ( marked in tables 1-3 and 5), the utility in calculating
the eigenvalues and eigenfunctions in one batch to sufficiently higher -
orders ( documented through the solution (28) of (1), with $b=2$), and the
applicability to a wide range of potentials. Provided that the potential
$V(q)$ gives rise to one minimum of $E_{k,l}^{(-2)}$ and an infinite
number of bound - states. Moreover, beyond its promise as being quite
handy ( on the computational and practical methodical sides), it offers
a useful perturbation prescription where the zeroth - order approximation
$\bar{l}^2E_{k,l}^{(-2)}$ inherits a substantial amount of the total energy.

The above has been a very limited review and a number of other useful
approaches such as those presented by Papp [9,32], Bender and Wu [33],etc,
have been left unattended. However, their accomplishments are indeed of
actual novelties.

Finally, the scope of PSLET applicability extends beyond the present
$D$ - dimensional spiked harmonic oscillator model. It could be applied
to angular momentum states of multi - electron atoms [34-36], relativistic
and non - relativistic quark - antiquark models [37], etc.
\newpage

\begin{center}
Appendix
\end{center}

Although some of the following expressions have appeared in previous
articles [5,20-22], we would like to repeat them to make this article self
contained.

Expansions about $x=0$ (i.e.$q=q_o$), yield\\
\begin{equation}
\frac{1}{q^{2}}=\sum^{\infty}_{n=0} (-1)^{n}~ \frac{(n+1)}{q_{o}^{2}}
~ x^{n}~\bar{l}^{-n/2},
\end{equation}\\
\begin{equation}
V(x(q))=\sum^{\infty}_{n=0}\left(\frac{d^{n}V(q_{o})}{dq_{o}^{n}}\right)
\frac{(q_{o}x)^{n}}{n!}~\bar{l}^{-n/2}.
\end{equation}\\
Equation (5) thus becomes\\
\begin{equation}
\left[-\frac{1}{2}\frac{d^{2}}{dx^{2}}+\frac{q_{o}^{2}}{\bar{l}}
\tilde{V}(x(q))\right]
\Psi_{k,l}(x)=\frac{q_{o}^2}{\bar{l}}E_{k,l}\Psi_{k,l}(x),
\end{equation}\\
with\\
\begin{eqnarray}
\frac{q_o^2}{\bar{l}}\tilde{V}(x(q))&=&q_o^2\bar{l}
\left[\frac{1}{2q_o^2}+\frac{V(q_o)}{Q}\right]
+\bar{l}^{1/2}B_1 x+ B_2 x^2+\frac{(2\beta+1)}{2} \nonumber\\
&+&(2\beta+1)\sum^{\infty}_{n=1}(-1)^n \frac{(n+1)}{2}x^n \bar{l}^{-n/2}
+\sum^{\infty}_{n=3}B_n x^n\bar{l}^{-(n-2)/2} \nonumber\\ &+&
\beta(\beta+1)\sum^{\infty}_{n=0}(-1)^n\frac{(n+1)}{2}x^n\bar{l}^{-(n+2)/2},
\end{eqnarray}\\
\begin{equation}
B_n=(-1)^n \frac{(n+1)}{2}
+\left(\frac{d^n V(q_o)}{dq_o^n}\right)\frac{q_o^{n+2}}{n! Q}.
\end{equation}\\
Equation (32), along with (33) and (34), is evidently the one - dimensional
Schr\"odinger equation for a perturbed harmonic oscillator\\
\begin{equation}
\left[-\frac{1}{2}\frac{d^2}{dx^2}+\frac{1}{2}w^2 x^2 +\varepsilon_o
+P(x)\right]X_{k}(x)=\lambda_{k}X_{k}(x),
\end{equation}\\
where $w^2=2B_2$,\\
\begin{equation}
\varepsilon_o =\bar{l}\left[\frac{1}{2}+\frac{q_o^2 V(q_o)}{Q}\right]
+\frac{2\beta+1}{2}+\frac{\beta(\beta+1)}{2\bar{l}},
\end{equation}\\
and $P(x)$ represents the remaining terms in eq.(33) as infinite power
series perturbations to the harmonic oscillator. One would then imply that\\
\begin{eqnarray}
\lambda_{k}&=&\bar{l}\left[\frac{1}{2}+\frac{q_o^2 V(q_o)}{Q}\right]
+\left[\frac{2\beta+1}{2}+(k+\frac{1}{2})w\right]\nonumber\\
&+&\frac{1}{\bar{l}}\left[\frac{\beta(\beta+1)}{2}+\lambda_{k}^{(0)}\right]
+\sum^{\infty}_{n=2}\lambda_{k}^{(n-1)}\bar{l}^{-n},
\end{eqnarray}\\
and\\
\begin{equation}
\lambda_{k} = q_o^2 \sum^{\infty}_{n=-2} E_{k,l}^{(n)}
\bar{l}^{-(n+1)}.
\end{equation}\\
Hence, equations (37) and (38) yield\\
\begin{equation}
E_{k,l}^{(-2)}=\frac{1}{2q_o^2}+\frac{V(q_o)}{Q}
\end{equation}\\
\begin{equation}
E_{k,l}^{(-1)}=\frac{1}{q_o^2}\left[\frac{2\beta+1}{2}
+(k +\frac{1}{2})w\right]
\end{equation}\\
\begin{equation}
E_{k,l}^{(0)}=\frac{1}{q_o^2}\left[ \frac{\beta(\beta+1)}{2}
+\lambda_{k}^{(0)}\right]
\end{equation}\\
\begin{equation}
E_{k,l}^{(n)}=\lambda_{k}^{(n)}/q_o^2  ~~;~~~~n \geq 1.
\end{equation}\\
Where $q_o$ is chosen to minimize $E_{k,l}^{(-2)}$, i. e.\\
\begin{equation}
\frac{dE_{k,l}^{(-2)}}{dq_o}=0~~~~
and~~~~\frac{d^2 E_{k,l}^{(-2)}}{dq_o^2}>0.
\end{equation}\\
Hereby, $V(q)$ is assumed to be well behaved so that $E_{k,l}^{(-2)}$ has
a minimum $q_o$ and there are well - defined bound - states.
Equation (43) in turn gives, with $\bar{l}=\sqrt{Q}$,\\
\begin{equation}
l_D-\beta=\sqrt{q_{o}^{3}V^{'}(q_{o})}.
\end{equation}\\
Consequently, the second term in Eq.(33) vanishes and the first term adds 
a constant to the energy eigenvalues. It should be noted that the energy term
$\bar{l}^2E_{k,l}^{(-2)}$  corresponds roughly to the energy of a classical
particle with angular momentum $L_z$=$\bar{l}$  executing circular motion of 
radius $q_o$ in the potential $V(q_o)$. It thus identifies the 
zeroth - order approximation, to all eigenvalues, as a classical 
approximation and the higher - order corrections as quantum fluctuations
around the minimum $q_o$, organized in inverse powers of $\bar{l}$.
The next correction to the energy series, $\bar{l}E_{k,l}^{(-1)}$,
consists of a constant term and the exact eigenvalues of the harmonic
oscillator $w^2x^2/2$.The shifting parameter
$\beta$ is determined by choosing
$\bar{l}E_{k,l}^{(-1)}$=0. This choice is physically motivated. In addition
to its vital role in removing the singularity at $l=0$, it also requires
the agreements between PSLET eigenvalues and eigenfunctions with
the exact well known ones
for the harmonic oscillator and Coulomb potentials.  Hence\\
\begin{equation}
\beta=-\left[\frac{1}{2}+(k+\frac{1}{2})w\right],
\end{equation}\\
where $w=\sqrt{3+q_o V^{''}(q_o)/V^{'}(q_o)}$, and primes of $V(q_o)$ denote
derivatives with respect to $q_o$. Then equation (33) reduces to\\
\begin{equation}
\frac{q_o^2}{\bar{l}}\tilde{V}(x(q))=
q_o^2\bar{l}\left[\frac{1}{2q_o^2}+\frac{V(q_o)}{Q}\right]+
\sum^{\infty}_{n=0} v^{(n)}(x) \bar{l}^{-n/2},
\end{equation}\\
where\\
\begin{equation}
v^{(0)}(x)=B_2 x^2 + \frac{2\beta+1}{2},
\end{equation}\\
\begin{equation}
v^{(1)}(x)=-(2\beta+1) x + B_3 x^3,
\end{equation}\\
and for $n \geq 2$\\
\begin{eqnarray}
v^{(n)}(x)&=&B_{n+2}~ x^{n+2}+(-1)^n~ (2\beta+1)~ \frac{(n+1)}{2}~ x^n
\nonumber\\
&+&(-1)^{n}~ \frac{\beta(\beta+1)}{2}~ (n-1)~ x^{(n-2)}.
\end{eqnarray}\\

\newpage
\begin{table}
\begin{center}
\caption{ 3D ground - state energies, in $\hbar=m=1$ units, for
$V(q)=(q^2+c_2/q^{5/2})/2$. Where $E_P$ represents PSLET results, Eq.(41),
and $\bar{l}^2E^{(-2)}$ is its zeroth - order approximation.
$E[4,4]$ shows the effect of the $P_{4}^{4}(1/\bar{l})$ Pad\'{e} approximant
,Eq.(42). $E_{VAM}$ from VAM, and $E_{DNI}$ from DNI [1a].}
\vspace{1cm}
\begin{tabular}{|cccccc|}
\hline\hline
$c_2$ & $\bar{l}^2E^{(-2)}$ & $E_P$ & $E[4,4]$ & $E_{VAM}$ & $E_{DNI}$\\
\hline
1000 & 44.003142 & 44.9554848 & 44.9554848 & 44.955485 & 44.955485\\
100  & 16.666664 & 17.541890 & 17.541890  & 17.541890 & 17.541890\\
10   & 7.00149 & 7.73515  & 7.73510  & 7.73511  & 7.73511\\
1    & 3.84771 & 4.31578  & 4.31413  & 4.32326  & 4.31731\\
0.1  & 3.11132 & 3.26984  & 3.26633  & 3.29602  & 3.26687\\
0.01 & 3.0116 & 3.0341   & 3.0344   & 3.0392   & 3.0367\\
0.001& 3.0012 & 3.0035   & 3.0040   & 3.0041  & 3.0040\\
\hline\hline
\end{tabular}
\end{center}
\end{table}
\newpage
\begin{table}
\begin{center}
\caption{ $D=2,\cdots,10$  ground - state energies, in $\hbar=m=1$ units,
 for the potential $V(q)=(q^2+10/q^{1.9})/2$.
Where $E_P$ represents PSLET results, Eq.(41),
and $\bar{l}^2E^{(-2)}$ is its zeroth - order approximation,
$E[4,4]$ shows the effect of the $P_{4}^{4}(1/\bar{l})$ Pad\'{e} approximant
, Eq.(42). $E_{VAM}$ from VAM, and $E_{DNI}$ from DNI [1a].}
\vspace{1cm}
\begin{tabular}{|cccccc|}
\hline\hline
$D$ & $\bar{l}^2E^{(-2)}$ & $E_P$ & $E[4,4]$ & $E_{VAM}$ & $E_{DNI}$\\
\hline
2 & 7.581 139 & 8.485 461 & 8.485 369 & 8.485 384  & 8.485 378\\
3 & 7.919 880 & 8.564 352 & 8.564 355 & 8.564 358 & 8.564 356\\
4 & 8.339 920 & 8.795 436 & 8.795 440 & 8.795 440 & 8.795 440\\
5 & 8.840 678 & 9.163 092 & 9.163 093 & 9.163 093 & 9.163 093\\
6 & 9.416 352 & 9.646 701 & 9.646 701 & 9.646 701 & 9.646 701\\
7 & 10.058 042 & 10.225 045& 10.225 045 & 10.225 045 & 10.225 045\\
8 & 10.755 870 & 10.879 077& 10.879 077 & 10.879 077 & 10.879 077\\
9 & 11.500 402 & 11.592 982& 11.592 982 & 11.592 982 & 11.592 982\\
10& 12.283 349 & 12.354 183& 12.354 183 & 12.354 183 & 12.354 183\\
\hline\hline
\end{tabular}
\end{center}
\end{table}
\newpage
\begin{table}
\begin{center}
\caption{ $2D$ - and $3D$ - nodeless states energies, with $l=0,\cdots,4$
(in $\hbar=m=1$ units), for the potential $V(q)=(q^2+1000/q^{b})/2$.
Where $E_{0,l}$ represents PSLET results with the $P_{4}^{4}(1/\bar{l})$
Pad\'{e} approximant, Eq.(42).}
\vspace{1cm}
\begin{tabular}{|ccccccc|}
\hline\hline
$D$ & $b$ & $E_{0,0}$ & $E_{0,1}$ & $E_{0,2}$ & $E_{0,3}$ & $E_{0,4}$\\
\hline
2 & 0.5 & 415.886751 & 415.898889 & 415.935293 & 415.995938 & 416.080780\\
  & 1   & 190.719321 & 190.735267 & 190.783089 & 190.862739 & 190.974135\\
  & 1.5 & 104.404517 & 104.427341 & 104.495769 & 104.609681 & 104.768874\\
  & 2   & 65.245553 & 65.277168 & 65.371918 & 65.529521 & 65.749510\\
  & 2.5 & 44.945030 & 44.986838 & 45.112071 & 45.320150 & 45.610129\\
  & 3   & 33.303511 & 33.356491 & 33.515080 & 33.778229 & 34.144222\\
3 & 0.5 & 415.889786& 415.914059& 415.962588& 416.035338& 416.132258\\
  & 1   & 190.72331 & 190.755196& 190.818940& 190.914475& 191.041704\\
  & 1.5 & 104.41022 & 104.455860& 104.547051& 104.683633& 104.865367\\
  & 2   & 65.253459 & 65.316665 & 65.442888 & 65.631753 & 65.882705\\
  & 2.5 & 44.95549 & 45.039054 & 45.205805 & 45.454976 & 45.785438\\
  & 3   & 33.31676 & 33.422634 & 33.633677 & 33.948503 & 34.365078\\
\hline\hline
\end{tabular}
\end{center}
\end{table}
\newpage
\begin{table}
\begin{center}
\caption{ $2D$  and $3D$  $k$ - state energies, in $\hbar=m=1$ units, for
the potential $V(q)=(q^2+1000/q^{3/2})/2$.
Where $E_P$ represents PSLET results, Eq.(41),
and $\bar{l}^2E^{(-2)}$ is its zeroth - order approximation,
$E[4,4]$ shows the effect of the $P_{4}^{4}(1/\bar{l})$ Pad\'{e}
approximant, Eq.(42).}
\vspace{1cm}
\begin{tabular}{|cccccc|}
\hline\hline
$D$ &$k$ & $l$ & $\bar{l}^2E^{(-2)}$ & $E_P$ & $E[4,4]$\\
\hline
2 & 1 & 0 & 105.40419 & 108.15083 & 108.15083\\
  &   & 1 & 105.67466 & 108.17379 & 108.17379\\
  &   & 2 & 105.96940 & 108.24263 & 108.24263\\
  &   & 3 & 106.28970 & 108.35721 & 108.35721\\
3 &   & 0 & 105.53648 & 108.15657 & 108.15657\\
  &   & 1 & 105.81892 & 108.20248 & 108.20248\\
  &   & 2 & 106.12628 & 108.29421 & 108.29421\\
  &   & 3 & 106.45983 & 108.43160 & 108.43160\\
2 & 2 & 0 & 107.3876 & 111.9017 & 111.9017\\
  &   & 1 & 107.7382 & 111.9248 & 111.9248\\
  &   & 2 & 108.1127 & 111.9940 & 111.9940\\
3 &   & 0 & 107.5600 & 111.9075 & 111.9074\\
  &   & 1 & 107.9224 & 111.9536 & 111.9536\\
  &   & 2 & 108.3092 & 112.0459 & 112.0459\\
\hline\hline
\end{tabular}
\end{center}
\end{table}
\newpage
\begin{table}
\begin{center}
\caption{ $k=2$ and $l=1,2$ energies, in $\hbar=m=1$ units, for the potential
$V(q)=(q^2+10/q^{2.1})/2$. Where $E_{2,l,P}$ represents PSLET results,
Eq.(41), $E_{2,1,V}$ from VAM, and $E_{2,1,ex}$ from DNI [1a].
$E_{2,l}[4,4]$ shows the effect of the $P_{4}^{4}(1/\bar{l})$ Pad\'{e}
approximant, Eq.(42).}
  
\vspace{1cm}
\begin{tabular}{|ccccccc|}
\hline\hline
$D$ & $E_{2,1,ex}$ & $E_{2,1,V}$ & $E_{2,1,P}$ & $E_{2,1}[4,4]$ &
$E_{2,2,P}$ & $E_{2,2}[4,4]$\\
\hline
2  & 16.543629& 16.543648& 16.541951& 16.543627& 17.380817& 17.381708\\
3  & 16.904445& 16.904446& 16.903172& 16.904444& 17.954856& 17.955444\\
4  & 17.381708& 17.381709& 17.380817& 17.381708& 18.606695& 18.607067\\
5  & 17.955444& 17.955446& 17.954856& 17.955444& 19.320461& 19.320691\\
6  & 18.607067& 18.607070& 18.606695& 18.607067& 20.083266& 20.083406\\
7  & 19.320691& 19.320693& 19.320461& 19.320691& 20.884936& 20.885021\\
8  & 20.083406& 20.083407& 20.083266& 20.083406& 21.717556& 21.717608\\
9  & 20.885021& 20.885022& 20.884936& 20.885021& 22.574996& 22.575027\\
10 & 21.717608& 21.717608& 21.717556& 21.717608& 23.452505& 23.452524\\
\hline\hline
\end{tabular}
\end{center}
\end{table}

\end{document}